\documentclass[11pt,a4paper]{article}
\pdfoutput=1
\usepackage{jheppub} 
\usepackage{amsfonts,amssymb,latexsym,amsmath,mathrsfs,amsbsy}
\usepackage{bm}
\usepackage{slashed,bbold}
\usepackage[hhmmss]{datetime}
\usepackage{dsfont}
\usepackage{feynmf}

\usepackage[latin1]{inputenc}

\title{Anomalous Strangeness Transport}

\author[a]{Eugenio Meg\'{\i}as,}
\author[b]{Miguel \'A. V\'azquez-Mozo}

\affiliation[a]{\sl Departamento de F\'{\i}sica At\'omica, Molecular y Nuclear  \\
\& Instituto Carlos I de F\'{\i}sica Te\'orica y Computacional, Universidad de Granada\\
Avda. de Fuente Nueva s/n, E-18071 Granada, Spain
}

\affiliation[b]{\sl Departamento de F\'{\i}sica Fundamental, Universidad de Salamanca \\ 
 Plaza de la Merced s/n,
 E-37008 Salamanca, Spain
  }

\emailAdd{emegias@ugr.es}
\emailAdd{vazquez@usal.es}

\abstract{
Nondissipative transport of strangeness is studied in a chiral hadronic plasma with three
flavors. In the phase in which chiral symmetry is preserved, strangeness transport is found 
to be driven by both an external magnetic field and fluid vorticity. As for the constitutive
relations of the baryon and electromagnetic currents, they exhibit 
vortical terms proportional to the strangeness chemical potential. 
In the superfluid phase, transverse nondissipative diffusion 
of the baryon, electromagnetic, and
strangeness charges is found, which survives in the limit of vanishing chiral imbalance
and mixes in a fashion similar to standard dissipative diffusion in
quark-gluon plasma.
}

\arxivnumber{}

\begin{document}

\maketitle

%\centerline{\bf Version: \today \,\,(\currenttime)}

\flushbottom

%%%%%%%%%%%%%%%%%%%%%%%%%%%%%%%%%%%%%%%%%%%%%%%%%%%%%%%%%
%%%%%%%%%%%%%%%%%%%%%%%%%%%%%%%%%%%%%%%%%%%%%%%%%%%%%%%%%

\section{Introduction and general setup}

Anomalous transport~\cite{Fukushima:2008xe,Son:2009tf,Sadofyev:2010pr,Neiman:2010zi,Landsteiner:2011cp,
Kirilin:2012mw} 
is known to be operative in a variety of physical setups 
where chiral imbalance is present
(see also~\cite{Fukushima:2012vr,Landsteiner:2016led} for reviews). Among them
is nuclear matter~\cite{Son:2004tq,Son:2007ny,Gatto:2011wc,Capasso:2013jva,
Fukushima:2019ugr,Fukushima:2021got,
Evans:2022hwr}
and in particular the dynamics
of the quark-gluon plasma experimentally
produced in heavy ion collisions~\cite{Kharzeev:2010gr,Hongo:2013cqa,Liao:2014ava,Kharzeev:2015znc,
Huang:2015oca,Li:2020dwr,Kharzeev:2022hqz,Grieninger:2023myf}, where
chirality imbalance may results from vacuum
configurations with nonzero topological number~\cite{Kharzeev:2001ev,Kharzeev:2007jp}. 
This points to the possibility of detecting
nondissipative charge transport induced either by a magnetic field or a nonzero fluid vorticity,
a current undertaking in experimental studies of heavy ion collisions~\cite{Tu:2019mso,
ALICE:2020siw,STAR:2021mii,
ALICE:2022ljz,STAR:2023gzg}.

Historically, strangeness production played a key role in the early attempts to find experimental signatures of
quark-gluon plasma production~\cite{Rafelski:2019twp}. Strangeness difussion, however, is not
disentangled from that of the electric and baryon charges, as 
a gradient in the density of one of them induces dissipative diffusion in the
others~\cite{Greif:2017byw}. 
In this short note, we explore {\em nondissipative} transport of strangeness in nuclear matter, modelized
as
a chiral hadronic fluid with three flavors in equilibrium and 
coupled to an external axial-vector gauge field 
affected by a 't Hooft anomaly, thus implementing chiral imbalance in the medium. 

The addition of a third flavor, 
besides making calculations 
more involved with respect to the~$N_{f}=2$ case studied in~\cite{Manes:2019fyw}, 
raises an important physical issue.
It concerns
the justification of the chiral approximation in the case of the strange quark, 
whose mass is similar to the effective temperature of the plasma,~$m_{s}/T\sim 1/3$. As
a result, chirality flip due to the strange quark mass would result 
in a dissipation of chirality in the sector with nonzero strangeness, effectively decoupling the 
strange quark from chiral transport which would be fully dominated by the two light flavors. 
It was argued however
in~\cite{Guo:2016nnq,Shi:2017cpu} that finite mass effects scale rather like $(m_{s}/T)^{2}$, 
bringing about a stronger suppression of chirality dissipation and making the three-flavor approximation 
a reliable starting point for the analysis of the system.   

Keeping this in mind, we study the constitutive relations of the strangeness 
current, as well as
the effect
of the strange flavor on other currents. Our model consists in a fermion fluid
coupled to Abelian vector and axial-vector external gauge fields denoted respectively by~$\mathcal{V}$ 
and~$\mathcal{A}$. In flavor space, the associated vector field one-forms are 
expanded in the basis spanned by 
the baryon number~($B$), electric charge~($Q$), and strangeness~($S$) matrices, given in terms of the
Gell-Mann matrices $\lambda_{a}$~($a=1,\ldots,8$) and the~$3\times 3$ identity matrix~$\mathbb{1}$ by
\begin{align}
B={1\over 3}\mathbb{1},
 \hspace*{1cm}
Q={e\over 2}\left(\lambda_{3}+{1\over \sqrt{3}}\lambda_{8}\right), \hspace*{1cm} 
S={1\over\sqrt{3}}\lambda_{8}-{1\over 3}\mathbb{1},
\end{align}
with~$e$ the elementary charge.
The isospin matrix~$I$, on the other hand, is obtained from 
the previous ones through the Gell-Mann--Nishijima 
relation~${1\over e}Q={1\over 2}(B+S)+I$. 

In addition to this, all form fields will be written using the electric-magnetic decomposition 
with respect to the fluid four-velocity~$u$. For the vector gauge field, we have
\begin{align}
\mathcal{V}=\boldsymbol{V}Q+iu\big(\mu_{B}B+\mu_{Q} Q+\mu_{S}S\big)
\equiv \boldsymbol{\mathcal{V}}_{M}+u\mathcal{V}_{E},
\label{eq:vector_gauge_field}
\end{align}
where~$\boldsymbol{V}$ is the magnetic piece of the 
electromagnetic one-form potential and $\mu_{B}$,~$\mu_{Q}$, and~$\mu_{S}$ are
the baryon number, charge, and strangeness chemical potentials respectively. Regarding the
axial-vector gauge field, on the other hand, we follow~\cite{Manes:2018llx,Manes:2019fyw}
and take it to be purely electric and proportional
to the identity matrix
\begin{align}
\mathcal{A}=iu\mu_{5}\mathbb{1}\equiv u\mathcal{A}_{E},
\label{eq:axial-vector_gauge_field}
\end{align}
with~$\mu_{5}$ the chiral chemical potential controlling chiral imbalance in the system. 
Moreover, the time component of the axial-vector gauge fields is taken
to be constant, which amounts to the condition
\begin{align}
\big(d+\mathfrak{a}\big)\mu_{5}=0,
\end{align}
where~$\mathfrak{a}\equiv\imath_{u}du$ is the fluid acceleration one-form, expressed here
in terms of the differential forms interior product~\cite{Nakahara:2003nw}.
Using all previous relations, the
field strengths of the vector and axial-vector gauge fields can be written as
\begin{align}
\mathcal{F}_{V}&=2i\mu_{B}\omega B+\mathbb{B}Q+2i\mu_{S}\omega S
-iuT\left[d\left({\mu_{B}\over T}\right)B
+d\left({\mu_{Q}\over T}\right)Q+d\left({\mu_{S}\over T}\right)S\right] \nonumber \\[0.2cm]
&\equiv \boldsymbol{F}_{V}+uE_{V}, \nonumber \\[0.2cm]
\mathcal{F}_{A}&=2i\mu_{5}\omega\mathbb{1}
\equiv \boldsymbol{F}_{A},
\end{align}
where~$\omega={1\over 2}(du+u\mathfrak{a})$ 
is the vorticity two-form,~$T$ is the equilibrium local temperature, 
and~$\mathbb{B}=d\boldsymbol{V}+2i\mu_{Q}\omega$ denotes the 
magnetic
field two-form\footnote{Here we follow the conventions of~\cite{Manes:2018llx,Manes:2019fyw}, with the
only exception that
our definition of the magnetic field includes the vorticity-dependent term~$2i\mu_{Q}\omega$
(see also~\cite{Jensen:2013kka,Manes:2020zdd}).}. 

After these preliminaries, we are ready to start discussing 
a hadronic fluid in the symmetric phase. The method to be employed here is the same one 
devised in ref.~\cite{Manes:2018llx},
and applied in~\cite{Manes:2019fyw} to the case of a two-flavor hadronic fluid. 
Our departing point is the Abelian Bardeen Chern-Simons form
\begin{align}
\omega^{0}_{5}(\mathcal{A},\mathcal{F}_{V},\mathcal{F}_{A})=-{i\over 4\pi^{2}}
{\rm Tr\,}\left[\mathcal{A}\left(\mathcal{F}_{V}^{2}
+{1\over 3}\mathcal{F}_{A}^{2}\right)\right],
\label{eq:CSnoGBNf=3_gen}
\end{align}
from where the equilibrium partition function~$W_{\rm eq}$ is computed by applying 
the Ma\~nes-Stora-Zumino transgression formula~\cite{Manes:1985df} to a one-parameter family of
connections interpolating between $\mathcal{V}=\boldsymbol{V}Q$, $\mathcal{A}=0$ and the configuration
given in eqs.~\eqref{eq:vector_gauge_field} and~\eqref{eq:axial-vector_gauge_field} 
(see~\cite{Manes:2018llx,Manes:2019fyw} for full details). 
The resulting partition function has the structure $W_{\rm eq}=W_{\rm bulk}+
W_{\rm bdy}$, where the first (nonlocal) piece defined on the
five-dimensional bulk~$\mathcal{M}_{5}$ has the form
\begin{align}
W_{\rm bulk}&=
-{i\over 4\pi^{2}}\int\limits_{\mathcal{M}_{5}}
u{\rm Tr\,}\bigg\{
\mathcal{A}_{E}\boldsymbol{F}_{A}^{2}+\mathcal{A}_{E}\boldsymbol{F}_{V}^{2}
+2\mathcal{V}_{E}\boldsymbol{F}_{A}\boldsymbol{F}_{V} 
\nonumber \\[0.2cm]
&\left.-2\Big[\boldsymbol{F}_{A}\big(\mathcal{A}_{E}^{2}+\mathcal{V}_{E}^{2}\big)
+2\boldsymbol{F}_{V}\mathcal{A}_{E}\mathcal{V}_{E}\Big]\omega
+{4\over 3}\mathcal{A}_{E}\Big(\mathcal{A}_{E}^{2}+3\mathcal{V}_{E}^{2}\Big)\omega^{2}\right\}.
\end{align}
The local term~$W_{\rm bdy}$, on the other hand, is defined on the boundary~$\partial\mathcal{M}_{5}$,
identified with the physical spacetime, and in our case it can be shown to be zero. This is a 
consequence of having chosen a purely electric 
axial-vector gauge field in eq.~\eqref{eq:axial-vector_gauge_field}. Notice that,
since in the following we are
going to be concerned only with the constitutive relations of vector currents, there is no 
problem in setting~$\boldsymbol{A}=0$ in the effective action before taking any variations.

\section{The strangeness covariant current}

The
vector and axial-vector (dual) covariant currents are computed by varying 
the bulk piece of the equilibrium partition function with respect to the field 
strengths~$\boldsymbol{F}_{V}$ and~$\boldsymbol{F}_{A}$ and keeping the boundary contributions~\cite{Jensen:2013kka,Manes:2018llx,Manes:2019fyw}. 
For the vector current, the result is
\begin{align}
\langle\star\boldsymbol{J}_{V}\rangle_{\rm cov}&={i\over 2\pi^{2}}u\Big(
\mathcal{A}_{E}\boldsymbol{F}_{V}+\boldsymbol{F}_{A}\mathcal{V}_{E}-2\mathcal{A}_{E}\mathcal{V}_{E}\omega\Big) \nonumber \\[0.2cm]
&=-{1\over 2\pi^{2}}\mu_{5}u\Big(
2i\mu_{B}\omega B+\mathbb{B}Q+2i\mu_{S}\omega S\Big).
\label{eq:starJV}
\end{align}
To compute the strangeness current from here, we take the trace of the
product of the vector current with the strangeness matrix,~$\langle \star\boldsymbol{J}_{S}\rangle_{\rm cov}
={\rm Tr\,}\big(S\langle\star\boldsymbol{J}_{V}\rangle_{\rm cov}\big)$. Since the dual vector current~\eqref{eq:starJV}
is purely electrical, taking a further Hodge dual leads to a four-vector whose 
covariant time component vanishes, 
$\langle J_{S,0}\rangle_{\rm cov}=0$, whereas the contravariant spatial components are given by
\begin{align}
\langle J_{S}^{i}\rangle_{\rm cov}&={N_{c}\over 6\pi^{2}}\mu_{5}\Big(
2\mu_{B}\omega^{i}+e\mathbb{B}^{i}-6\mu_{S}\omega^{i}\Big),
\label{eq:JSunbrokencs}
\end{align}
with~$N_{c}$ the number of colors. In addition,~$\mathbb{B}^{i}$ and~$\omega^{i}$ are
expressed in terms of the components of the two-forms~$\mathbb{B}$
and~$\omega$ introduced above by
\begin{align}
\mathbb{B}^{i}&={1\over 2}\epsilon^{ijk}\mathbb{B}_{jk}, \nonumber \\[0.2cm]
\omega^{i}&={1\over 2}\epsilon^{ijk}\omega_{jk}.
\end{align}
The result for the strangeness current in
the symmetric phase shows the existence of chiral nondissipative transport of strangeness charge driven by 
both an external magnetic field and fluid vorticity. 

To gain some further physical insight about the different terms in~\eqref{eq:JSunbrokencs}, 
we have to take into account that 
the strange quark carries both baryon 
and electric charge, so strangeness transport is entangled with the transport of
the other two charges. From a formal viewpoint, this results from 
the nonorthogonality of the~$\{B,Q,S\}$
flavor basis.
Going back to eq.~\eqref{eq:starJV},
we notice that the term containing the magnetic field comes from
the contribution proportional to~${\rm Tr\,}(QS\mathbb{1})$,
whose origin is a triangle
diagram with one axial-vector, one electromagnetic, and one strangeness current insertion. 
Nondissipative transport of strangeness associated with this term 
is therefore consequence of the standard chiral magnetic effect affecting the strange quarks.
Something similar happens with the vortical baryonic contribution,
stemming from a term proportional to~${\rm Tr\,}\big(SB\mathbb{1}\big)$. Its diagrammatic origin lies 
in a triangle with one axial-vector, one baryonic, and one strangeness currents. Here again, the chiral vortical
effect for the baryonic current (cf. the results of ref.~\cite{Manes:2019fyw}) induces strangeness transport due
to the nonzero baryon number of the strange quark. The upshot of all previous considerations is  
that the only ``genuine'' anomalous strangeness transport
comes from the vortical term in~\eqref{eq:JSunbrokencs} weighted by the
strangeness chemical potential~$\mu_{S}$, 
originated in a triangle diagram with one axial-vector and two strange current insertions, 
whose flavor factor is~${\rm Tr\,}\big(S^{2}\mathbb{1}\big)$.

The Bardeen-Zumino (BZ) currents, on the other hand, can be evaluated from the Chern-Simons 
form~\eqref{eq:CSnoGBNf=3_gen} using the explicit expressions given
in ref.~\cite{Manes:2018llx}. In particular, we find that in our case the BZ
vector current is identical to the corresponding covariant current
\begin{align}
\langle\star\boldsymbol{J}_{V}\rangle_{\rm BZ}
&=\langle\star\boldsymbol{J}_{V}\rangle_{\rm cov},
\label{eq:JVBZ_general_QBS}
\end{align}
in agreement with the fact that the boundary partition function is zero and so are all
consistent vector currents\footnote{The consistent axial-vector current, however, is nonzero, since
the magnetic part of the axial-vector gauge field in the boundary partition function
cannot be set to zero prior to taking variations with respect to~$\mathcal{A}$.}.  

Let us now turn to the analysis of the system after chiral symmetry breaking. 
The computation of the covariant currents in this phase 
can be carried out by the appropriate transformation of the BZ currents
in the symmetric phase using the Nambu-Goldstone boson matrix~$U$, as shown in eqs.~(6.26) and~(6.29) of
ref.~\cite{Manes:2018llx}.
This matrix is parametrized 
in terms of the pion, kaon, and~$\eta_{8}$-meson fields by
\begin{align}
U=\exp\left[{i\sqrt{2}\over f_{\pi}}\left(
\begin{array}{ccc}
{1\over\sqrt{2}}\pi^{0}+{1\over\sqrt{6}}\eta_{8} & \pi^{+} & K^{+}
\\
\pi^{-} & -{1\over\sqrt{2}}\pi^{0}+{1\over\sqrt{6}}\eta_{8} & K^{0}
\\
K^{-} & \overline{K}^{0} & -\sqrt{{2\over 3}}\eta_{8}
\end{array}
\right)\right],
\label{eq:matrixU3flvs}
\end{align}
where~$f_{\pi}\approx 92\,\mbox{MeV}$ is the pion decay constant. Unlike in the unbroken phase studied earlier 
where the (dual) currents 
were purely electrical,
now the vector and axial-vector currents have both electric and magnetic components. Here we are ultimately
interested in the contravariant 
spatial components, so we only need to evaluate the electric part of the corresponding
three-form currents, the magnetic parts giving the covariant time 
components upon taking the Hodge dual. The calculation
is long and involved but follows the same steps detailed in ref.~\cite{Manes:2019fyw}. It
leads to the following result for the covariant strangeness current
\begin{align}
\langle J^{i}_{S}\rangle_{\rm cov}&=
{N_{c}\over 6\sqrt{3}\pi^{2}f_{\pi}}T\epsilon^{ijk}\partial_{j}\eta_{8}\partial_{k}\left({\mu_{B}\over T}\right)
-{eN_{c}\over 6\sqrt{3}\pi^{2}f_{\pi}}T\epsilon^{ijk}\partial_{j}\eta_{8}\partial_{k}\left({\mu_{Q}\over T}\right)
\nonumber \\[0.2cm]
&-{N_{c}\over 2\sqrt{3}\pi^{2}f_{\pi}}T\epsilon^{ijk}\partial_{j}\eta_{8}\partial_{k}\left({\mu_{S}\over T}\right)
+{N_{c}\over 6\pi^{2}f_{\pi}^{2}}\mu_{5}\epsilon^{ijk}\Big[i\partial_{j}K^{+}\partial_{k}K^{-}
+e\mathbb{V}_{j}\partial_{k}\big(K^{+}K^{-}\big)\Big]
\nonumber \\[0.2cm]
&-{eN_{c}\over 3\pi^{2}f_{\pi}^{2}}\mu_{5}K^{+}K^{-}\big(e\mathbb{B}^{i}+\mu_{Q}\omega^{i}\big)
+{iN_{c}\over 6\pi^{2}f_{\pi}^{2}}\mu_{5}\epsilon^{ijk}\partial_{j}K^{0}\partial_{k}\overline{K}^{0}
\label{eq:strangeness_current_all_final} \\[0.2cm]
&+{N_{c}\over 3\pi^{2}f_{\pi}^{2}}\mu_{5}\mu_{S}\Big(
K^{+}K^{-}+K^{0}\overline{K}^{0}\Big)\omega^{i}
+{N_{c}\over 6\pi^{2}}\mu_{5}\Big(e\mathbb{B}^{i}
+2\mu_{B}\omega^{i}-6\mu_{S}\omega^{i}\Big),
\nonumber
\end{align}
where~$\mathbb{V}_{i}$ are the components of the electromagnetic potential and we have dropped
terms of third order and higher in the meson fields.
Notice that, despite the explicit appearance of the electromagnetic potential, the previous
expression remains gauge invariant. 

To check that the anomalous strangeness transport encoded in the constitutive 
relations~\eqref{eq:JSunbrokencs} and~\eqref{eq:strangeness_current_all_final}
is indeed a nondissipative phenomenon, we look at whether the different 
transport coefficients remain invariant under 
the time reversal operation~$\mathsf{T}$. From the transformation of 
the classical Nambu-Goldstone matrix field~\eqref{eq:matrixU3flvs},~$\mathsf{T}:U\rightarrow U^{\dagger}$
(see, for example,~\cite{Invitation2QFT}), we find~$\mathsf{T}:(\pi^{0},\eta_{8})\rightarrow
(-\pi^{0},-\eta_{8})$, $\mathsf{T}:(\pi^{\pm},K^{\pm})\longrightarrow (-\pi^{\mp},-K^{\mp})$,
and~$\mathsf{T}:(K^{0},\overline{K}^{0})\rightarrow (-\overline{K}^{0},-K^{0})$, whereas the
vorticity and the gauge and magnetic fields
satisfy~$\mathsf{T}:(\omega^{i},\mathbb{V}_{i},\mathbb{B}^{i})
\rightarrow (-\omega^{i},-\mathbb{V}_{i},-\mathbb{B}^{i})$.
Combining these transformations with 
the one for the strangeness current,~$\mathsf{T}:J^{i}_{S}\rightarrow -J^{i}_{S}$, we conclude
that all transport coefficients in 
eqs.~\eqref{eq:JSunbrokencs} and~\eqref{eq:strangeness_current_all_final} 
are~$\mathsf{T}$-even\footnote{These same transformations, 
together with the~$\mathsf{T}$-odd character of the electromagnetic
and baryonic currents, imply as well the nondissipative character
of the anomalous transport phenomena to be analyzed in the next section, both in the 
symmetric and the broken phases (cf.
the discussion of the two-flavor case in ref.~\cite{Manes:2019fyw}).}. Notice
that all chemical potentials are invariant under time reversal, 
since they are proportional to the time component of the corresponding background gauge fields.

The first three terms in~\eqref{eq:strangeness_current_all_final}, 
driven by the gradients of the chemical potentials,
can be interpreted as describing nondissipative
strangeness diffusion. Unlike
the dissipative case studied in~\cite{Greif:2017byw} in which the current points 
along the gradient, this is mediated
by the gradient of the $\mathsf{T}$-odd $\eta_{8}$-meson
and is itself normal to the charge gradient.   
Interestingly, these are the only contributions 
surviving in the absence of chiral imbalance.
In fact, all terms in~\eqref{eq:strangeness_current_all_final}, 
apart form the ones depending on~$\mu_{S}$ 
have to be interpreted as resulting from the mixing between strangeness~$S$ and
electric charge~$Q$ and baryon number~$B$. Notice however the conspicuous absence of terms 
depending on the meson fields and proportional to the 
baryon chemical potential~$\mu_{B}$, indicating that baryon anomalous transport only
contributes to the strangeness covariant current through the meson-independent BZ terms.

\section{Other currents}

A nonvanishing strangeness chemical potential~$\mu_{S}$ 
also has effects on the constitutive relations of other currents. 
In the unbroken phase, an explicit evaluation of the electromagnetic 
current~$\langle \boldsymbol{J}_{\rm em}\rangle_{\rm cov}
={\rm Tr\,}\big(Q\langle\boldsymbol{J}_{V}\rangle_{\rm cov}\big)$
shows the existence of a chiral vortical effect mediated by~$\mu_{5}\mu_{S}$
\begin{align}
\langle J_{\rm em}^{i}\rangle_{\rm cov}&={eN_{c}\over 3\pi^{2}}\mu_{5}\Big(
e\mathbb{B}^{i}-\mu_{S}\omega^{i}\Big).
\label{eq:em_current_cov_symphase}
\end{align}
While the first, chiral magnetic effect 
term has its source in the standard triangle diagram with one axial-vector and two
electromagnetic currents, the 
second one has the same diagrammatic origin as the first term in eq.~\eqref{eq:JSunbrokencs}.
A further peculiarity of~$N_{f}=3$ is that~${\rm Tr\,}Q=0$, which removes from the 
constitutive relations of the electromagnetic covariant
current
a vortical term proportional to the baryon number chemical potential~$\mu_{B}$, that is however present
for~$N_{f}=2$ (notice that this contribution was not explicitly computed in 
ref.~\cite{Manes:2019fyw}, where the baryon chemical potential was
set to zero from the start). 

In the broken phase, the electromagnetic and baryon currents are obtained along the same lines
as the strangeness current shown in eq.~\eqref{eq:JSunbrokencs}. For the first one, we 
find
\begin{align}
\langle J^{i}_{\rm em}\rangle_{\rm cov}&=
{eN_{c}\over 12\pi^{2} f_{\pi}}T\epsilon^{ijk}\left(\partial_{j}\pi^{0}+{1\over\sqrt{3}}\partial_{j}\eta_{8}\right)\partial_{k}\left({\mu_{B}\over T}\right) \nonumber \\[0.2cm]
&+{e^{2}N_{c}\over 12\pi^{2} f_{\pi}}T\epsilon^{ijk}\left(\partial_{j}\pi^{0}+{1\over\sqrt{3}}\partial_{j}\eta_{8}\right)\partial_{k}\left({\mu_{Q}\over T}\right)
\nonumber \\[0.2cm]
&-{eN_{c}\over 6\sqrt{3}\pi^{2}f_{\pi}}T\epsilon^{ijk}\partial_{j}\eta_{8}\partial_{k}\left({\mu_{S}\over T}\right)
-{eN_{c}\over 3\pi^{2}f_{\pi}^{2}}\mu_{5}
\Big(\pi^{+}\pi^{-}+K^{+}K^{-}\Big)\big(e\mathbb{B}^{i}+\mu_{Q}\omega^{i}\big)
\nonumber \\[0.2cm]
&+{eN_{c}\over 6\pi^{2}f_{\pi}^{2}}\mu_{5}\epsilon^{ijk}
\Big[i\Big(\partial_{j}\pi^{+}\partial_{k}\pi^{-}+\partial_{j}K^{+}\partial_{k}K^{-}\Big)
+e\mathbb{V}_{j}\partial_{k}\Big(\pi^{+}\pi^{-}+K^{+}K^{-}\Big)\Big] 
\label{eq:emcurrents_finalresultNf=3} \\[0.2cm]
&+{eN_{c}\over 3\pi^{2}f_{\pi}^{2}}\mu_{5}\mu_{S}K^{+}K^{-}\omega^{i}
+{eN_{c}\over 3\pi^{2}}\mu_{5}\Big(e\mathbb{B}^{i}-\mu_{S}\omega^{i}\Big).
\nonumber
\end{align}
Again, the first three terms proportional to the chemical potentials gradients
give rise to a nondissipative transverse diffusion of electric charge, similar to the corresponding
effect spotted in the constitutive relations for the 
strangeness covariant current~\eqref{eq:strangeness_current_all_final}. 
As for the remaining contributions, only the terms proportional to~$\mu_{5}\mu_{S}$
represent electric charge transport induced by the anomalous transport of other conserved charges,
in this case strangeness (here, as a consequence of~${\rm Tr\,}Q=0$, 
there are no contributions resulting from the mixing between electric and 
baryonic charges). 
Finally, the meson-independent term is the BZ electromagnetic current that, as we explained above,
coincides in our model with the covariant electomagnetic current in the symmetric 
phase given in eq.~\eqref{eq:em_current_cov_symphase}. 

We complete our analysis with the calculation of the constitutive relations for the baryonic 
covariant current. In the symmetric phase, the result is
\begin{align}
\langle J^{i}_{\rm bar}\rangle_{\rm cov}&=
-{N_{c}\over 3\pi^{2}}\mu_{5}\big(\mu_{B}-\mu_{S}\big)\omega^{i}.
\label{eq:baryoncurrents_finalresult_symmetricNf=3}
\end{align}
Here we find a vortical effect similar to the one encountered in 
the electromagnetic current~\eqref{eq:em_current_cov_symphase}
proportional to~$\mu_{5}\mu_{S}$, this time in combination with the one driven by
a nonvanishing baryonic chemical potential.
Thus, a nonzero strangeness chemical potential gives rise to 
a vortical term proportional to~${\rm Tr\,}(B\mathbb{1}S)$, originating in a triangle with 
an axial-vector, a baryonic, and a strangeness current.
Once more, the identity~${\rm Tr\,}Q=0$ eliminates any electromagnetic contribution to the
anomalous transport of the baryonic charge.  

Chiral symmetry breaking adds
transverse nondissipative diffusion terms to the constitutive relation for 
the baryonic current, driven by the gradients of the electric charge and strangeness
chemical potentials 
\begin{align}
\langle J^{i}_{\rm bar}\rangle_{\rm cov}&=
{eN_{c}\over 12\pi^{2}f_{\pi}}T\epsilon^{ijk}\left(\partial_{j}\pi^{0}
+{1\over \sqrt{3}}\partial_{j}\eta_{8}\right)\partial_{k}\left({\mu_{Q}\over T}\right)
+{N_{c}\over 6\sqrt{3}\pi^{2}f_{\pi}}T\epsilon^{ijk}\partial_{j}\eta_{8}
\partial_{k}\left({\mu_{S}\over T}\right)\nonumber \\[0.2cm]
&-{N_{c}\over 3\pi^{2}}\mu_{5}\big(\mu_{B}-\mu_{S}\big)\omega^{i}.
\label{eq:baryoncurrents_finalresultNf=3}
\end{align}
Notice the absence in this case of meson-dependent contributions proportional to the chiral chemical
potential~$\mu_{5}$.

%{\color{red}
%As in the case of strangeness studied in the previous section, anomalous
%baryon number and electric charge transport 
%are also nondissipative phenomena. Taking into account that 
%both the electromagnetic and baryon currents change sign under time reversal,
%and using the transformations listed in page~\ref{pag:T-transf}, it can be checked
%that all transport coefficients in the constitutive
%relations~\eqref{eq:em_current_cov_symphase}-\eqref{eq:baryoncurrents_finalresultNf=3} 
%are invariant under~$\mathsf{T}$ 
%(cf. the discussion of the two-flavor case in ref.~\cite{Manes:2019fyw}).
%In the superfluid phase this implies that 
%the terms surviving the~$\mu_{5}\rightarrow 0$ limit can be also interpreted 
%as describing nondissipative diffusion, similar to the one found in the strangeness
%current. 
%}

\section{Closing remarks}

We studied the anomalous transport of strangeness in a chiral hadronic fluid with three flavors 
at equilibrium, analyzing the cases where chiral symmetry is
preserved and spontaneously broken. In the symmetric case, 
our main conclusion is that there are nondissipative
mechanisms of strangeness transport driven by vorticity and an external magnetic field.  

After chiral spontaneous symmetry breaking, we found a number of contributions to the constitutive relations
mediated by the meson fields and/or their gradients.  
More remarkably, we showed how all three 
currents~\eqref{eq:strangeness_current_all_final},~\eqref{eq:emcurrents_finalresultNf=3}, 
and~\eqref{eq:baryoncurrents_finalresultNf=3} contain terms that survive the limit of vanishing chiral imbalance ($\mu_{5}\rightarrow 0$),
depending on the gradients of the three chemical potential. This can be interpreted as 
describing transverse nondissipative diffusion in the chiral hadronic fluid (i.e., normal to the
direction set by the charge gradient). As a matter of fact, 
the contributions  mentioned 
can be written in a way very much reminiscent of the structure found in~\cite{Greif:2017byw}
for dissipative diffusion
\begin{align}
\left.\left(
\begin{array}{c}
\langle\vec{J}_{\rm bar}\rangle_{\rm cov} \\
\langle\vec{J}_{\rm em}\rangle_{\rm cov} \\
\langle\vec{J}_{S}\rangle_{\rm cov}
\end{array}
\right)\right|_{\mu_{5}=0}
=
\left(
\begin{array}{ccc}
\vec{\kappa}_{BB} & \vec{\kappa}_{BQ} &  \vec{\kappa}_{BS} \\
\vec{\kappa}_{QB} & \vec{\kappa}_{QQ} & \vec{\kappa}_{QS} \\
\vec{\kappa}_{SB} & \vec{\kappa}_{SQ} & \vec{\kappa}_{SS}
\end{array}
\right)\times
\left(
\begin{array}{c}
\vec{\nabla}\left({\mu_{B}\over T}\right) \\
\vec{\nabla}\left({\mu_{Q}\over T}\right) \\
\vec{\nabla}\left({\mu_{S}\over T}\right)
\end{array}
\right),
\label{eq:matrix_gradients}
\end{align}
where the notation indicates the matrix product should be carried out using the three-dimensional 
vector cross-product between the corresponding entries. 
The calculations presented above show that most entries in the
(vectorial) diffusion coefficient matrix are nonzero and determined by the gradients
of the~$\pi^{0}$ and~$\eta_{8}$ meson fields, namely
\begin{align}
\vec{\kappa}_{BB}&=\vec{0}, \nonumber \\[0.2cm]
\vec{\kappa}_{QQ}&=e\vec{\kappa}_{QB}=e\vec{\kappa}_{BQ}
={e^{2}N_{c}\over 12\pi^{2}f_{\pi}}T\vec{\nabla}\left(\pi^{0}
+{1\over\sqrt{3}}\eta_{8}\right),  \\[0.2cm]
e\vec{\kappa}_{SS}&=3\vec{\kappa}_{QS}=3\vec{\kappa}_{SQ}=-3e\vec{\kappa}_{BS}=-3e\vec{\kappa}_{SB}
=-{eN_{c}\over 2\sqrt{3}\pi^{2}f_{\pi}}T\vec{\nabla}\eta_{8}.
\nonumber
\end{align} 
We see that, as in the dissipative case,
anomaly-mediated diffusion mixes the different gradients among themselves. This is again a consequence
of the use of a nonorthogonal basis of conserved charges, which implies
microscopic degrees of freedom carry all three quantum numbers. Notice moreover that all terms
in~\eqref{eq:matrix_gradients} have their origin in the axial anomaly responsible for the
electromagnetic decays of neutral Nambu-Goldstone 
mesons~$\pi^{0}\rightarrow 2\gamma$ and~$\eta\rightarrow 2\gamma$.

Since this mixed nondissipative transport of conserved charges is at work even in the absence of chirality 
imbalance, it would be interesting to 
explore this phenomenon in more precise modelizations of the quark-gluon plasma in order to decide whether
they can be detected in current heavy-ion collision
facilities. 
To the extend that the model used here provides a reliable 
description of the physics of quark-gluon plasma produced in 
heavy ion collisions, our results might point to a novel 
way of searching for nondissipative phenomena in these physical
systems by focusing on strangeness transport. 

\section*{Acknowledgments} 

We thank Juan~L.~Ma\~nes and Manuel Valle for discussions and collaboration on related topics. 
The work of E.M. is supported by the project PID2020-114767GB-I00 and by the Ram\'on y Cajal Program under Grant RYC-2016-20678 funded by MCIN/AEI/10.13039/ 501100011033 and by ``FSE Investing in your future'', by the FEDER/Junta de Andaluc\'{\i}a-Consejer\'{\i}a de Econom\'{\i}a y Conocimiento 2014-2020 Operational Programme under Grant A-FQM-178-UGR18, by Junta de Andaluc\'{\i}a under Grant FQM-225, and by the ``Pr\'orrogas de Contratos Ram\'on y Cajal'' Program of the University of Granada. M.A.V.-M. acknowledges the financial support from the
Spanish Science Ministry through research grant PID2021-123703NB-C22
(MCIU/AEI/FEDER, EU), as well as from Basque Government grant
IT1628-22.

\bibliographystyle{bibliostyle}
\bibliography{biblio_file}

\end{document}